\documentclass[showpacs,preprint]{revtex4}
\usepackage{amsfonts}
\usepackage{amsmath}
\usepackage{amssymb}
\usepackage{graphicx}
\usepackage{color}
\usepackage{epsfig}
\usepackage{remreset}
\usepackage{hyperref}

\begin{document}

\title{Gauss-Bonnet correction to Holographic thermalization: two-point
functions, circular Wilson loops and entanglement entropy}
\author{Yong-Zhuang Li$^{1}$}
\author{Shao-Feng Wu$^{1,2}$\footnote{%
Corresponding author. Email: sfwu@shu.edu.cn; Phone: +86-021-66136202.}}
\author{Guo-Hong Yang$^{1,2}$}
\affiliation{$^{1}$Department of physics, Shanghai University, Shanghai, 200444, P. R.
China}
\affiliation{$^{2}$The Shanghai Key Lab of Astrophysics, Shanghai, 200234, P. R. China}

\begin{abstract}
We study the thermalization of a class of 4-dimensional strongly coupled
theories dual to a 5-dimensional AdS-Vaidya spacetime with Gauss-Bonnet
curvature corrections. We probe the thermalization using the two-point
functions, the expectation values of circular Wilson loops and entanglement
entropy. When boundary separation is small, we observe that the
thermalization times of these observables have the weak dependence on the
Gauss-Bonnet coupling constant $\alpha $. In addition, the growth rate of
entanglement entropy density is nearly volume-independent. We also show that
a new kind of swallow-tail behavior may exhibit in the thermalization of the
two-point function when $\alpha $ is negative and $\ell$ is large enough. At
large negative $\alpha $ ($\alpha \lesssim -0.1$) the relationship between
the critical thermalization time of entanglement entropy and the boundary
separation encounters certain \textquotedblleft phase
transition\textquotedblright .
\end{abstract}

\maketitle

\section{Introduction}

The AdS/CFT correspondence \cite{Maldacena9711200} has been fruitful in
revealing universal features of strongly coupled field theories by
gravitational description. The duality has entered into the regime of
far-from-equilibrium physics. In particular, the so-called \textquotedblleft
holographic thermalization\textquotedblright\ has attracted many interest,
which is well motivated by the demand of describing the fast thermalization
of the quark gluon plasma produced in heavy ion collisions at the
Relativistic Heavy Ion Collider \cite{MullerReview,Muller11} and of
describing some condensed matter systems prepared under quantum quenches
\cite{Polkovnikov}.

In this note, we will investigate the thermalization process of a class of
strongly coupled conformal field theories (CFTs) by its dual gravity with
Gauss-Bonnet (GB) curvature corrections. Generally, the higher derivative
corrections appear in any quantum gravity theory (like string theory) from
quantum or stringy effect. These corrections may be holographic dual to $1/N$
or $1/\lambda $ corrections in some gauge theories. In terms of the standard
correspondence between type IIB string theory on $AdS_{5}\times S^{5}$ and
the $\mathcal{N}$=4 super Yang-Mills theory (SYM) theory, the stringy effect
gives the leading order $1/\lambda $ corrections, which has the form of $%
\alpha ^{\prime 3}R^{4}$, rather than $R^{2}$. However, the quadratic
curvature corrections may also arise in the framework of string theory.
Consider the dual between the $\mathcal{N}=2$ $Sp(N)$ gauge theory (with 4
fundamental and 1 antisymmetric traceless hypermultiplets) and type IIB
string theory on $AdS_{5}\times X^{5}$ with $X^{5}\simeq S^{5}/\mathbb{Z}%
_{2} $ \cite{Fayyazuddin1}. The curvature squared term arises in the
effective action on the D7 and O7 world-volumes \cite{Aharony2} and leads to
certain subleading $1/N$ corrections. Note that the corrections allow
independent values of the two central charges $a$ and $c$ of the dual field
theory, in contrast to the $\mathcal{N}$=4 SYM where $a=c$. This feature is
essential for the violation of the well-known Kovtun-Starinets-Son (KSS)
bound of the ratio of the shear viscosity $\eta $ to the entropy density $s$
\cite{5,GBalpha,Liualpha}. Beside its relation with string theory mentioned
above, there are also some motivations to involve the GB curvature
corrections{\footnote{%
Note that any other curvature squared term can be reduced to GB term by
field redefinitions and disregarding six or more derivatives \cite{Liualpha}.%
}} in holographic theories, two of which are particular for the holographic
thermalization. i) Keeping in mind the vastness of the string landscape, one
could not rule out certain situations in which GB curvature is dominated in
all the higher derivative corrections \cite{Liualpha}. ii) Comparing with
the other higher curvature theory of gravity, the equations of motion of GB
gravity are only second order in derivatives. This makes the holographic
calculations being easily dealt with. iii) It was found that GB gravity
captures the causality constraints \cite{GBalpha,Liualpha} which can be also
inferred from the positivity of the energy in CFT analysis \cite{Hofman1005}%
. Such fundamental correspondence conveys a piece of strong evidence
supporting the AdS/CFT conjecture. iv) Since the KSS bound is not effective
in theories dual to GB gravity, it promotes to treat GB correction as a
dangerous source of violation for the features\ which is universal in usual
holographic models \cite{Liu1005}. For instance, the universality of the
ratio of gap frequency over critical temperature on holographic
superconductors breaks in the presence of the GB coupling \cite{Gregory1005}%
. v) The exact black hole solution of GB gravity, particularly the Vaidya
type solution, have been constructed \cite{GBVaidya}. This paves the way for
our study on holographic thermalization. vi) Entanglement entropy (EE) is an
important non-local observable of holographic thermalization. However, there
is not yet a general holographic prescription for computing EE in higher
curvature theory of gravity (such as the one with $R^{4}$ correction from
type IIB string theory), except in the case of GB gravity \cite%
{GB1,Myers1101,Sinha1305}.

We will study three non-local observables, including the two-point
correlation function, the (circular) Wilson loop expectation values and the
EE (with circular entangling surfaces). They can be evaluated in the saddle
point approximation in terms of some extremal geometric objects in the bulk.
We will adapt the simple AdS-Vaidya model \cite%
{Arrastia10,Aparicio11,Johnson11,Bala11,Keranen11,Galante1205,Hubeny1302,Ours,stricker1307,QC,GBHT}%
, which describes a homogeneous{\footnote{%
Recently the inhomogeneous holographic thermalization has been studied in
\cite{inhomo1307}.}} falling thin shell of null dust and is a good
quantitative approximation of the background generated by the perturbation
of a time-dependent scalar field \cite{Bhattacharyya0904} and of the model
of Ref. \cite{Lin08}. In term of Vaidya models, some interesting features of
holographic thermalization have been found. For instance, i) the
thermalization process is top-down, i.e. the high momentum modes thermalize
faster than long wavelength ones. This is contrary to the thermalization
behavior of the weak field theory \cite{Son01} but is very natural in the
Vaidya models, where the collapsing shell from the boundary passes the
extremal surface of the smaller probe more faster than the one of the larger
probe. ii) EE thermalizes slowest among the probes and sets the time-scale
for thermalization. This feature may also be understood intuitively since
other observables are sensitive only to a subset of all degrees of freedom
of the field theory which are involved in the calculation of entropy \cite%
{Muller11}. iii) It was observed that the evolution of holographic EE
includes an intermediate stage during which it is a simple linear function
of time. The linear regime is not obvious when the boundary separation is
small, but it will be when the boundary separation increases. This result
matches well with the behavior seen in d=2 CFTs \cite{Cardy05,Cardy07}. Also
it is consistent with the evolution of coarse-grained entropy in nonlinear
dynamical systems. In particular, in classical 4-dim SU(2) lattice gauge
theory, the linear growth rate of coarse-grained entropy, that is generally
described by the Kolmogorov-Sina\"{\i} entropy rate \cite{Latora99}, is
shown to be an extensive quantity \cite{Muller00}. For strongly coupled
field theory with gravity dual, it has been found \cite{Bala11} that the
growth rate of EE density in d=2 CFTs is also nearly volume-independent for
small boundary volumes. For large volumes, however, the growth rate of EE
approaches a constant limit\footnote{%
In Ref. \cite{Bala11}, the maximal growth rate is used to characterize the
linear growth, since the linear regime covers the time at which the growth
rate is maximal.}. Recently, the large volume behavior of EE has been
studied in very general situations \cite{Maldacena1303,Liu1005,Ours}.

We will focus on studying how the thermalization time of various probes and
the growth of EE at small volume are affected by the GB curvature
corrections. It is interesting that we find out both the universal features
which are independent with the curvature corrections and the new
thermalization behavior that is induced by the curvature corrections.

It should be noted that the higher curvature corrections to holographic
thermalization have been also discussed using the quasi-static approximation
\cite{HC1,HC2} and the Vaidya model \cite{Liu1005,Ours,QC,GBHT},
respectively. In particular, Ref. \cite{QC} investigated the thermalization
time scale in SYM plasmas with the leading type IIB string theory
corrections. It was found that both $\alpha ^{\prime }$ and $1/N$
contributions decrease just a very little the thermalization time of UV
modes. We will also show the weak effects\ of GB correction on the
thermalization time when the boundary separation is small. Moreover, the GB
correction to the two-point function and the rectangular Wilson loop has
been studied in Ref. \cite{GBHT}. We will compare our results with theirs at
the last section.

\section{The GB\ gravity}

Consider the 5D GB gravity with a negative cosmological constant $\Lambda $
\cite{GBmetric},%
\begin{equation}
I=\frac{1}{16\pi G_{N}^{(5)}}\int d^{5}x\sqrt{-g}\Big[-2\Lambda +\mathcal{R}+%
\frac{\alpha }{2}L^{2}\mathcal{L}_{GB}\Big],  \label{Eq1}
\end{equation}%
where

\begin{equation}
\Lambda =-\frac{6}{L^{2}},\ \mathcal{L}_{GB}=\mathcal{R}^{2}-4\mathcal{R}%
_{\mu \nu }\mathcal{R}^{\mu \nu }+\mathcal{R}_{\mu \nu \rho \sigma }\mathcal{%
R}^{\mu \nu \rho \sigma },
\end{equation}%
$L$ is AdS spacetime radius and $\alpha $ is GB coupling constant. It should
be noted that there exists a constraint $-\frac{7}{36}\leq \alpha \leq \frac{%
9}{100}$ by respecting the causality of dual field theory on the boundary
\cite{Liualpha,GBalpha} or preserving the positivity of the energy flux in
CFT analysis \cite{Hofman1005}. However, we still calculate three probes
with $-\frac{7}{36}\leq \alpha \leq \frac{1}{4}$. As a result, one could
extract some information on how the violation of the causality constraint of
$\alpha $\ influences the thermalization.

For simplicity we will set $L=1$ hereafter. The action gives the equation of
gravitational field
\begin{equation}
\mathcal{R}_{\mu \nu }-\frac{1}{2}g_{\mu \nu }\mathcal{R}=-\Lambda g_{\mu
\nu }+\frac{\alpha }{2}\mathcal{H}_{\mu \nu },  \label{Eq2}
\end{equation}%
where
\begin{equation}
\mathcal{H}_{\mu \nu }=\frac{1}{2}g_{\mu \nu }\mathcal{L}_{GB}-2\mathcal{R}%
\mathcal{R}_{\mu \nu }+4\mathcal{R}_{\mu \gamma }\mathcal{R}^{\gamma
}{}_{\nu }+4\mathcal{R}_{\gamma \delta }\mathcal{R}^{\gamma }{}_{\mu
}{}^{\delta }{}_{\nu }-2\mathcal{R}_{\mu \gamma \delta \lambda }\mathcal{R}%
_{\nu }{}^{\gamma \delta \lambda }.
\end{equation}%
Eq. (\ref{Eq2}) has a static black brane solution \cite{GBmetric}
\begin{equation}
ds^{2}=-r^{2}f(r)dt^{2}+\frac{1}{r^{2}f(r)}dr^{2}+\frac{r^{2}}{L_{AdS}^{2}}d%
\mathbf{x}^{2},  \label{Eq3}
\end{equation}%
where $f(r)=\frac{1}{2\alpha }\Big[1-\sqrt{1-4\alpha \Big(1-\frac{r_{h}^{4}}{%
r^{4}}\Big)}\Big]$, $r_{h}$ is event horizon radius, and the effective AdS
radius is given by $L_{AdS}^{2}=\frac{1+\sqrt{1-4\alpha }}{2}$. $\mathbf{x}%
=(x_{1},...,x_{d-1})$ correspond to the spatial coordinates on the boundary.

To model the thermalization processes of strongly coupled field theories by
a homogeneous falling thin shell of null dust in AdS spacetime, we invoke a
Vaidya type solution \cite{GBVaidya}:
\begin{equation}
ds^{2}=\frac{1}{z^{2}}[-f(z,v)dv^{2}-2dzdv+\frac{1}{L_{AdS}^{2}}d\mathbf{x}%
^{2}],  \label{Eq4}
\end{equation}%
where $f(z,v)=\frac{1}{2\alpha }\left[ 1-\sqrt{1-4\alpha (1-m(v)z^{4})}%
\right] $ and $z$ is the inverse of radial coordinate $r$. The mass function
is
\begin{equation}
m(v)=\frac{M}{2}\left[ 1+\tanh \left( \frac{v}{v_{0}}\right) \right] ,
\label{massfun}
\end{equation}%
where $M$ denotes the mass of the black brane for $v>v_{0}$ and $v_{0}$
represents a finite shell thickness. We will be interested in the zero
thickness limit, which means to set the energy deposition on the boundary as
instantaneous. But for later numerical convention, we set $v_{0}=0.01$.
Moreover, we will set $M=1$ and the meaning will be interpreted below.

One may notice that there are various forms of the GB black brane metric in
references\footnote{%
For instance, see Refs. \cite{Gregory1005,Liualpha,Kuang}.}, which are
related by a simple coordinate transformation, namely the effective AdS
radius may disappear or appear in certain component of metric. The
difference does not change the dimensionless quantities, such as $\eta /s$
in natural units. However, one should be careful to deal with the
dimensional quantities like the temperature, boundary separation and
thermalization time. We select the form of Eq. (\ref{Eq3}) because of two
reasons. One is that it is asymptotically AdS which respects the same
symmetry of the CFTs. The other is that the Hawking temperature $T=\frac{%
r_{h}}{\pi }=\frac{M^{1/4}}{\pi }$ is independent with $\alpha $ if we fix $%
M=1$. Consider that we use the collapse of a massless shell to model a
certain thermal quench, or a sudden injection of energy due to a heavy-ion
collision, and this energy will be involved in the mass $M$ ultimately \cite%
{QC}. Thus, if we adapt the metric Eq. (\ref{Eq3}) and fix $M=1$, we have
fixed the injected energy and the ultimate equilibrium temperature
simultaneously, which makes the comparison of thermalization processes with
different $\alpha $ more reasonably. Note that these two good situations do
not happen if one does not select the form of the metric suitably. Also see
Ref. \cite{QC}, in which $M$ can be fixed but $T$ will be dependent with the
't Hooft coupling, vice versa.

\section{The non-local observables}

In order to explore the dynamics and the scale dependence of thermalization
processes, the local observables can not provide sufficient information.
Here we will study the equal-time two-point function, Wilson loop
expectation values and EE. Using the AdS/CFT correspondence, the former two
can be calculated by the space-like geodesic and the extremal surface which
extend in the bulk and end on the boundary, respectively. The EE in the
field theory dual to GB gravity can also be obtained by extremizing some
geometric quantities in the bulk but is a little more complicated, since the
holographic formula of EE in GB gravity has the nontrivial correction (not
same as the Wald entropy) to the one of Einstein gravity \cite%
{Fursaev06,GB1,Myers1101,Sinha1305}.

\subsection{The two-point function}

Consider the space-like geodesic connecting two boundary points $%
(t,x_{1})=(t_{0},-\ell /2)$ and $(t,x_{1})=(t_{0},\ell /2)$. We have set
that all other spatial directions are identical at the two end points. One
can parameterize such a geodesic by $v=v(x),$ $z=z(x)$ where we have denoted
$x_{1}=x$. The length of geodesic is given by extremizing the action%
\begin{equation}
\mathcal{L}(\ell ,t_{0})=\int_{-\ell /2}^{\ell /2}dx\frac{1}{z}\sqrt{\frac{1%
}{L_{AdS}^{2}}-2z^{\prime }v^{\prime }-f(z,v)v^{\prime 2}},  \label{Geo}
\end{equation}%
where the prime indicates derivative with respect to $x$. To obtain the
minimal length of the geodesic, we need to solve the two equations of motion
that are derived from Eq. (\ref{Geo}):
\begin{equation}
2L_{AdS}^{2}zz^{\prime \prime 2}+\{L_{AdS}^{2}zv^{\prime }(v^{\prime }\frac{%
\partial {f}}{\partial {v}}+2z^{\prime }\frac{\partial {f}}{\partial {z}}%
)-2f^{2}L_{AdS}^{2}v^{\prime 2}+f[2+L_{AdS}^{2}v^{\prime }(zv^{\prime }\frac{%
\partial {f}}{\partial {z}}-4z^{\prime })]\}=0,  \label{36}
\end{equation}%
\begin{equation}
2zv^{\prime \prime }-\{\frac{2}{L_{AdS}^{2}}+v^{\prime }[v^{\prime }(z\frac{%
\partial {f}}{\partial {z}}-2f)-4z^{\prime }]\}=0.  \label{37}
\end{equation}%
Keeping in mind with the symmetry of the geodesic, one can impose the
boundary conditions%
\begin{equation}
z(0)=z_{\ast },\;v^{\prime }(0)=v_{\ast },\;z^{\prime }(0)=0,\;v^{\prime
}(0)=0.  \label{boun1}
\end{equation}%
The two parameters $z_{\ast }$ and $v_{\ast }$ are determined by the
constrain equations%
\begin{equation}
z(\pm \frac{\ell }{2})=z_{0},\;v(\pm \frac{\ell }{2})=t_{0},  \label{boun2}
\end{equation}%
where $z_{0}$ is an IR bulk cut-off.

\begin{figure}[tbp]
\centering
\includegraphics[width=0.24\textwidth]{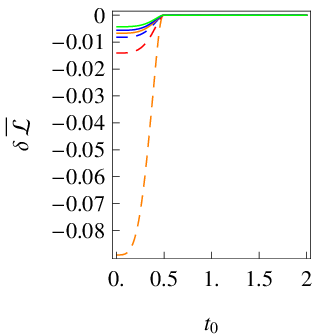} \includegraphics[width=0.24%
\textwidth]{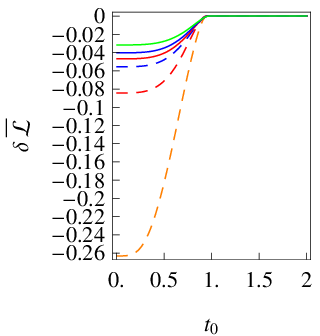} \includegraphics[width=0.24\textwidth]{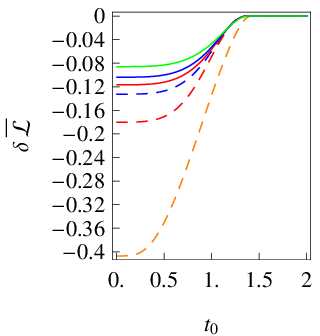} %
\includegraphics[width=0.23\textwidth]{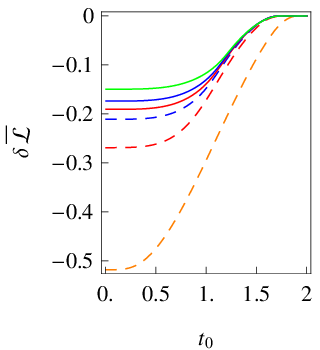}
\caption{$\protect\delta \bar{\mathcal{L}}$ as a function of $t_{0}$ for
boundary scale $\ell =1,2,3,4$ (from left to right) with different GB
coupling constants $\protect\alpha =-0.19,-0.1,-0.05,0,0.1,0.24$ according
to the green, blue, red, blue-dashed, red-dashed, orange-dashed lines,
respectively. For simplicity we will adapt the same identification among
curves and GB couplings from this figure to Fig. \protect\ref{timeandrate}.}
\label{twopoint1}
\end{figure}

Using Eq. (\ref{boun1}) and Eq. (\ref{boun2}), we can numerically solve Eqs.
(\ref{36}) and (\ref{37}) to calculate the minimal length of the geodesic by
Eq. (\ref{Geo}). We plot $\delta \bar{\mathcal{L}}(\ell ,t_{0})=\frac{\delta
\mathcal{L}(\ell ,t_{0})-\delta \mathcal{L}_{thermal}(\ell )}{\ell }$ as a
function of the boundary time $t_{0}$, the boundary separation $\ell $ and
GB coupling $\alpha $ in Fig. \ref{twopoint1}. The regularized length $%
\delta \mathcal{L}(\ell ,t_{0})$ is given by subtracting the cut-off
dependent part $\mathcal{L}_{div}=-2L_{AdS}\log {z_{0}}$ from Eq. (\ref{Geo}%
). Note that $\mathcal{L}_{div}$ can be obtained in the vacuum background.
In addition, the subscript \textquotedblleft $thermal$\textquotedblright\
denotes the regularized thermal equilibrium value that can be calculated in
the background of pure black branes.

One key quantity to characterize the thermalization process is the
thermalization time. In general, the real thermalization time $\tau _{real}$
should be defined as the time when the observable reaches the thermal value.
There are also other definitions of thermalization time which can describe
certain aspects of the thermalization process \cite{Bala11}. i) The
half-thermalization time, at which the observable reaches half of their
equilibrium value; ii) the rapid thermalization time, at which
thermalization proceeds most rapidly; iii) the critical thermalization time $%
\tau _{crit}$, at which the tip of the extremal line or surface grazes the
middle of the shell at $v=0$. Beside the real thermalization time that can
be read from Fig. \ref{twopoint1}, we also have interested in the critical
time with the mind of its two features. First, $\tau _{crit}$ can be
determined by the black brane geometry outside the infalling shell:%
\begin{equation}
\tau _{crit}=\int_{z_{0}}^{z_{\ast }}\frac{dz}{f(z)},  \label{tcrit}
\end{equation}%
where $z_{\ast }$ depends on the boundary separation and $f(z)=\frac{1}{%
2\alpha }[1-\sqrt{1-4\alpha (1-Mz^{4})}]$. One can find that the more
complicated dynamical Vaidya geometry is not necessary to compute $\tau
_{crit}$. Consequently, the behavior of $\tau _{crit}$ can be analytically
analyzed in certain extent, as we will show later. Second, it has been
pointed out in \cite{Keranen11} that $\tau _{crit}$ provides a lower bound
of $\tau _{real}$ in general\footnote{%
This can be understood as follows. At first, for smaller times $t_{0}$ the
geodesics (or extremal surfaces) necessarily pass through the shell and the
probes are time dependent. Second, $\tau _{crit}$ can not necessarily fix $%
\tau _{real}$, because even at later times there may be a geodesic which
penetrates the shell and has a shorter length than the one that only extends
in the region outside the shell. In short, when the shell's instantaneous
position is at $z_{\ast }$, the geodesic may still pass through the shell.
So it will need more time to reach thermal equilibrium. As an example, by
studying the holographic thermalization in asymptotically Lifshitz
spacetimes, the authors of Ref. \cite{Keranen11} have found that for
dynamical exponent $\tilde{z}=1$, $\tau _{real}$ agrees with $\tau _{crit}$
in the parameter space they scanned. But for other value of $\tilde{z}$, $%
\tau _{real}$ is indeed larger than $\tau _{crit}$ for sufficiently large $%
\ell $.}. Moreover, we have checked numerically that $\tau _{crit}\simeq
\tau _{real}$ for three probes when the boundary separation is small enough.
Thus one may use $\tau _{crit}$ to suspect or understand some properties of $%
\tau _{real}$.

\begin{figure}[tbp]
\centering
\includegraphics[width=0.44\textwidth]{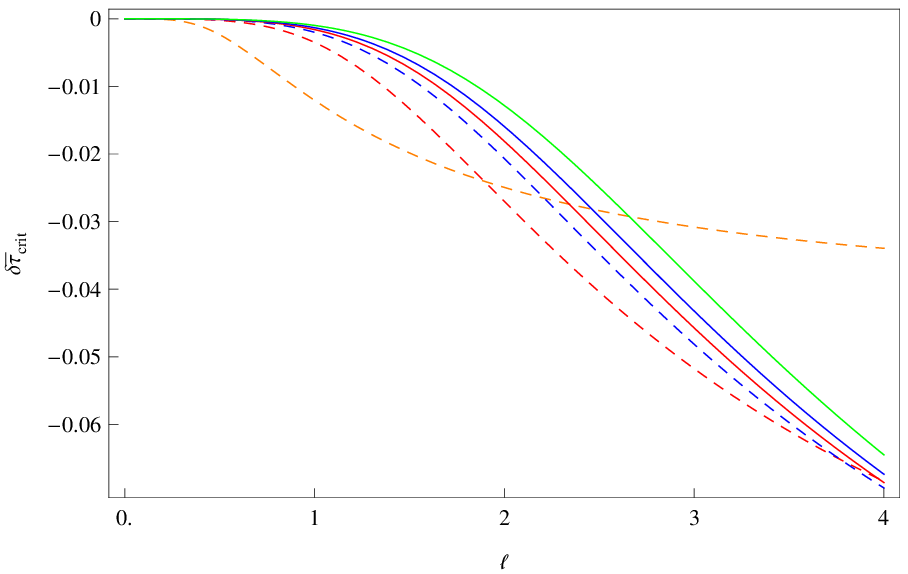} %
\includegraphics[width=0.43\textwidth]{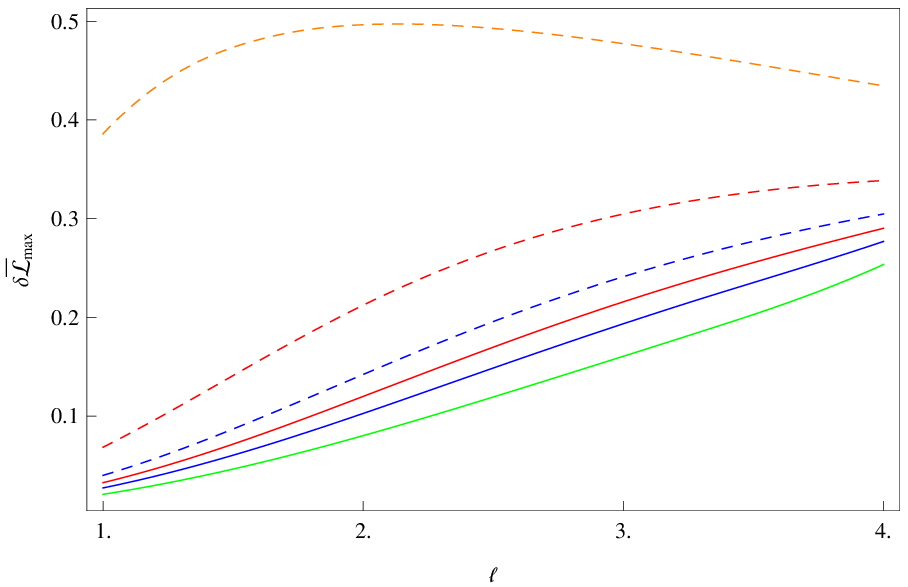}
\caption{(Left) The relative critical thermalization time $\protect\delta
\bar{\protect\tau}_{crit}(\equiv \frac{\protect\tau _{crit}-(\ell /2)}{\ell
/2})$ as a function of spatial scale $\ell $ with different $\protect\alpha $%
. (Right) Maximal growth rate of two-point function ($\protect\delta \bar{%
\mathcal{L}}_{max}$) as a function of boundary scalae $\ell $.}
\label{timeandratetp}
\end{figure}

From Fig. \ref{twopoint1}, one can find a nontrivial result that $\tau
_{real}$ is nearly independent with the GB curvature effect when $\ell $ is
small, particularly for the cases with $\alpha \neq 0.24$ (which exceeds far
more than the causality bound $-\frac{7}{36}\leq \alpha \leq \frac{9}{100}$%
). This behavior also appears in the critical thermalization time. To be
more clear, we plot the relative critical thermalization time $\delta \bar{%
\tau}_{crit}(\equiv \frac{\tau _{crit}-(\ell /2)}{\ell /2})$ as a function
of boundary separation $\ell $\ in the left panel of Fig. \ref{timeandratetp}%
. Note that the right panel about the maximal growth rate is plotted for
later comparison.

Beside the small dependence of critical thermalization time on the $\alpha $%
, we also find two interesting results. i) The critical thermalization time
has the non-monotonic dependence on the GB coupling constant $\alpha $. When
the boundary separation is small, the critical thermalization time decreases
as $\alpha $ increases. But it seems that there is the opposite behavior
when the boundary separation is large enough. One can see the tendency more
clearly for the case of Wilson loop in next subsection. It should be noted
that our numerical precision of $\tau _{real}$ is not high enough when $\ell
$ is large, so we can not identify the non-monotonic feature from $\tau
_{real}$. But the behavior of $\delta \bar{\tau}_{crit}$ gives a hint that $%
\tau _{real}$ could have the non-monotonic feature. ii) For the small
boundary separation the relative critical thermalization time is nearly
vanishing. In other word, the critical thermalization time saturates the
causality bound $\tau _{crit}\sim \frac{\ell }{2}$ \cite{Arrastia10}. We
note that this feature can be understood easily, since the geodesic with
small boundary separation in the static black brane will only extend near
the boundary. There, the metric is asymptotically AdS and can be written as
\begin{equation}
ds^{2}=\frac{1}{L_{AdS}^{2}z^{2}}\Big[-dv^{2}-2L_{AdS}^{2}dvdz+d\mathbf{x}%
^{2}\Big].  \label{flat}
\end{equation}%
For the two-point function one can calculate $z_{\ast }$ using the static
metric (\ref{flat}) and solving the equation of motion analytically. Thus,
we have
\begin{equation}
z(x)=\frac{\sqrt{L_{AdS}^{4}z_{\ast }^{2}-x^{2}}}{L_{AdS}^{2}},\ z_{\ast }=%
\frac{R_{\ast }}{L_{AdS}^{2}}  \label{eq1}
\end{equation}%
where $R_{\ast }=\ell /2$. Consequently, the critical thermalization time
can be given by%
\begin{equation}
\tau _{crit}=\int_{z_{0}}^{z_{\ast }}L_{AdS}^{2}dz\varpropto
L_{AdS}^{2}z_{\ast }=R_{\ast }=\ell /2.  \label{eq2}
\end{equation}

\subsection{The Wilson loop}

\begin{figure}[tbp]
\centering
\includegraphics[width=0.24\textwidth]{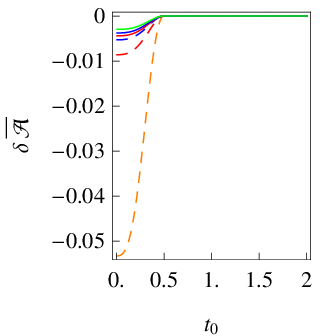} %
\includegraphics[width=0.24\textwidth]{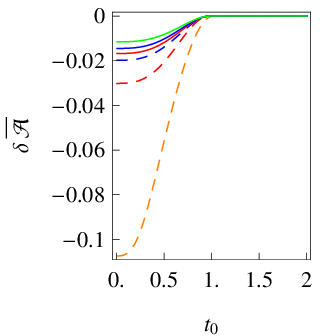} \includegraphics[width=0.24%
\textwidth]{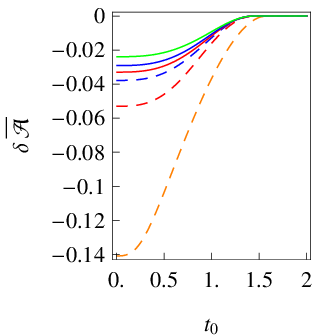} \includegraphics[width=0.24\textwidth]{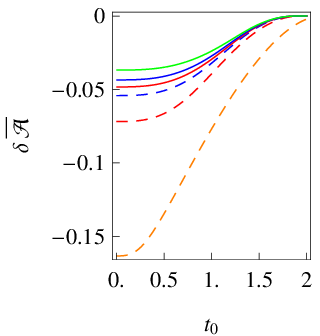}
\caption{$\protect\delta \bar{\mathcal{A}}$ as a function of $t_{0}$ for
circular Wilson loop radii $R$=0.5,~1,~1.5,~2 (from left to right) with
different $\protect\alpha $. }
\label{wilson1}
\end{figure}

For the circular Wilson loop one can choose a 2-dimensional plane $%
\{x_{1},x_{2}\}$ on the boundary in which this loop is described using the
polar coordinate $\{\rho ,\phi \}$. Then the minimal area surface can be
represented by $z(\rho )$ and $v(\rho )$ with respect to the azimuthal
symmetry in the $\phi $-direction. The area function is given by
\begin{equation}
\mathcal{A}(R,t_{0})=\int_{0}^{R}d\rho \frac{\rho }{L_{AdS}z(\rho )^{2}}%
Q(\rho )  \label{SNG}
\end{equation}%
where $Q(\rho )=\sqrt{\frac{1}{L_{AdS}^{2}}-2z^{\prime }v^{\prime
}-f(z,v)v^{\prime 2}}$ and $^{\prime }\equiv {\frac{d}{d\rho }}$. The
equations of motion are given by%
\begin{equation*}
2L_{AdS}^{2}\rho zz^{\prime \prime }-4L_{AdS}^{2}\rho f^{2}v^{\prime
2}-L_{AdS}^{2}z\left[ 4L_{AdS}^{2}v^{\prime }z^{\prime 2}-\rho v^{\prime 2}%
\frac{\partial {f}}{\partial {v}}-2z^{\prime }(1+\rho v^{\prime }\frac{%
\partial {f}}{\partial {z}})\right]
\end{equation*}%
\begin{equation}
-fL_{AdS}^{2}v^{\prime }\left[ 2z^{\prime }(4\rho +L_{AdS}^{2}zv^{\prime
})-\rho zv^{\prime }\frac{\partial {f}}{\partial {z}}\right] +4f\rho =0,
\label{wl19}
\end{equation}%
\begin{equation}
L_{AdS}^{2}\rho zv^{\prime \prime }-\frac{1}{2}L_{AdS}^{2}\rho zv^{\prime 2}%
\frac{\partial {f}}{\partial {z}}+(2\rho -L_{AdS}^{2}zv^{\prime
})[L_{AdS}^{2}v^{\prime }(fv^{\prime }+2z^{\prime })-1]=0.  \label{wl20}
\end{equation}%
To avoid a numerical issue at $\rho =0$, we solve Eq. (\ref{wl19}) and Eq. (%
\ref{wl20}) in the neighborhood of the midpoint by expanding around $\rho =0$
to quadratic order
\begin{equation}
z(\rho )=z_{\ast }-\frac{f(z_{\ast },v_{\ast })}{2L_{AdS}^{2}z_{\ast }}\rho
^{2},\ v(\rho )=v_{\ast }+\frac{\rho ^{2}}{2z_{\ast }L_{AdS}^{2}}.
\label{4344}
\end{equation}%
Eq. (\ref{4344}) will be used as the boundary conditions. Thus, we can plot $%
\delta \bar{\mathcal{A}}=\frac{\delta \mathcal{A}(R,t_{0})-\delta \mathcal{A}%
_{thermal}(R)}{\pi R^{2}}$ as a function of the boundary time $t_{0}$ for
different circular Wilson loop radius $R$ and GB coupling constant $\alpha $
in Fig. \ref{wilson1}. Note that we have regulated the minimal area by
subtracting the cut-off dependent piece:
\begin{equation}
\delta \mathcal{A}(R,t_{0})=\mathcal{A}(R,t_{0})-\mathcal{A}_{div}(R)
\label{45}
\end{equation}%
where $\mathcal{A}_{div}(R)=\frac{R}{z_{0}}$.

\begin{figure}[tbp]
\centering
\includegraphics[width=0.44\textwidth]{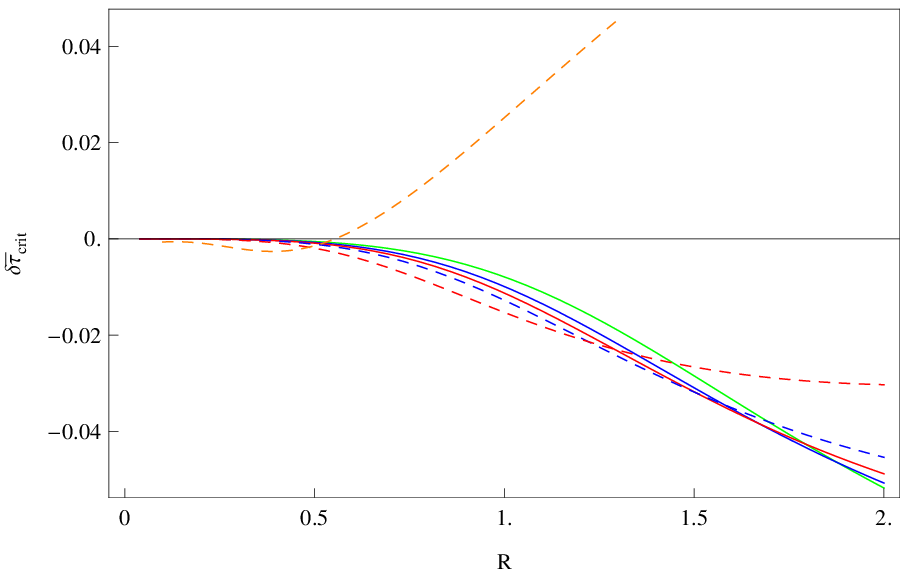} %
\includegraphics[width=0.43\textwidth]{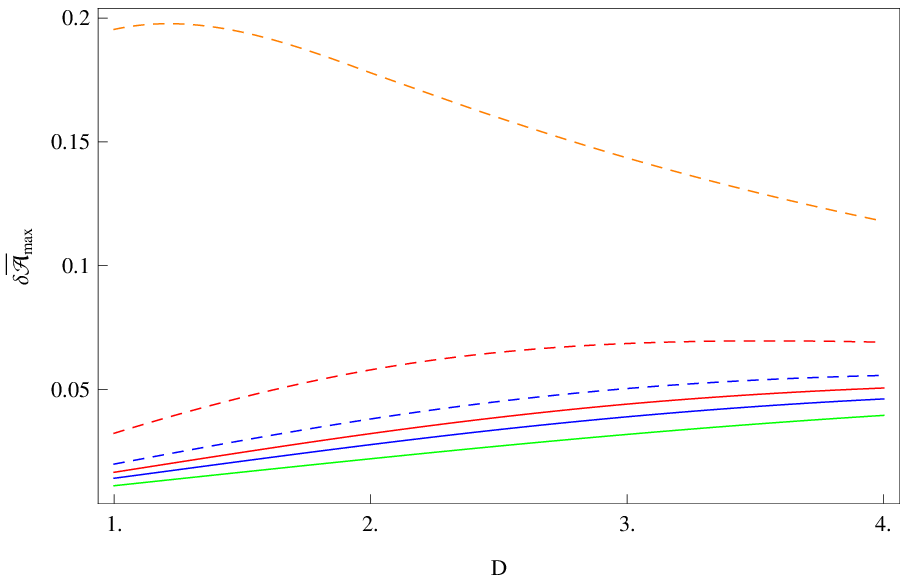}
\caption{(Left) The relative critical thermalization time $\protect\delta
\bar{\protect\tau}_{crit}$ as a function of spatial scale $R$ with different
$\protect\alpha $. (Right) Maximal growth rate of Wilson loop density ($%
\protect\delta\bar{\mathcal{A}}_{max}$) as a function of diameter of
entangled region.}
\label{timeandratewl}
\end{figure}

Repeating the analysis performed for two-point functions, we plot the
relative critical thermalization time $\delta \bar{\tau}_{crit}(\equiv \frac{%
\tau _{crit}-R}{R})$ and the maximal growth rate of minimum area density $%
\delta \bar{\mathcal{A}}_{max}$ in Fig. \ref{timeandratewl}. The pictures
show the similar behavior with the two-point function. Note that one can
check Eq. (\ref{eq1}) and (\ref{eq2}) holding for the case of Wilson loops.
In this case $R_{\ast }=R$.

\subsection{Holographic EE}

It has been proposed that the EE of a spatial region $V$ in a $d$%
-dimensional CFT with gravity dual is given by \cite{HEE3}
\begin{equation}
S(V)=\frac{1}{4G_{N}^{d+1}}\int_{\Sigma }\sqrt{h}  \label{entang}
\end{equation}%
where $\Sigma $ is the minimal surface whose boundary coincides with the
boundary of the region $V$ and $h$ corresponds to the determinant of the
induced metric of the minimal surface. In 5D GB gravity, Eq. (\ref{entang})
should be generalized to be the form of \cite{GB1,Myers1101}
\begin{equation*}
S(V)=\frac{1}{4G_{N}^{5}}\int_{\Sigma }dx^{3}\sqrt{h}(1+\alpha L^{2}\mathcal{%
R}_{\Sigma })+\frac{\alpha L^{2}}{2G_{N}^{5}}\int_{\partial \Sigma }dx^{2}%
\sqrt{\sigma }\mathcal{K}
\end{equation*}%
where $\mathcal{R}_{\Sigma }$ is the induced scalar curvature of $\Sigma $. $%
\sigma $ is the determinant of the induced metric on the boundary of $\Sigma
$. $\mathcal{K}$ is the trace of its extrinsic curvature. In fact, the last
term is the Gibbons-Hawking term which provide a good variational principle
in extremizing this functional.

Defining a unit normal vector of $\partial \Sigma $ in $\Sigma $ by $n^{a}$,
one can evaluate the extrinsic curvature
\begin{equation}
\mathcal{K}=\sigma ^{ab}\nabla _{a}n_{b}|_{z=z_{0}}  \label{27}
\end{equation}%
The induced metric on $\Sigma $ is given by%
\begin{equation}
ds_{\Sigma }^{2}=\frac{Q^{2}(\rho )}{z^{2}}d\rho ^{2}+\frac{1}{%
z^{2}L_{AdS}^{2}}d\Omega ^{2}  \label{28}
\end{equation}%
Then the normal unit vector is given by $n_{a}=\sqrt{h_{\rho \rho }}\delta
_{\rho a}$. And the volume function is given by
\begin{equation}
\mathcal{V}(R,t_{0})=4\pi \int_{0}^{R}\Big[\frac{\rho ^{2}}{z^{3}L_{AdS}^{2}}%
Q(\rho )+q(\rho )\Big]d\rho ^{2}  \label{46}
\end{equation}%
where
\begin{equation*}
q(\rho )=\frac{2\alpha }{z^{3}L_{AdS}^{2}Q(\rho )}%
\{z^{2}[2-f(z,v)L_{AdS}^{2}v^{\prime 2}]-2z(\rho +zL_{AdS}^{2}v^{\prime
})z^{\prime }+\rho ^{2}z^{\prime 2}\}.
\end{equation*}

Again, we will use the method in Section III.B by expanding the equations of
motion (They can be given by Eq. (\ref{46}) but we will not give out the
cumbersome result clearly.) at a point $\rho _{p}$ near the midpoint to
quadratic order. Then one gets the boundary conditions
\begin{equation}
z(\rho )=z_{\ast }-\frac{f(z_{\ast },v_{\ast })}{2L_{AdS}^{2}z_{\ast }}\rho
^{2},\ v(\rho )=v_{\ast }+\frac{\rho ^{2}}{2z_{\ast }L_{AdS}^{2}}.
\label{bb47}
\end{equation}%
For our aim, we subtract the volume functional Eq. (\ref{46}) with the
cut-off dependent piece
\begin{equation}
\delta \mathcal{V}(R,t_{0})=\mathcal{V}(R,t_{0})-\mathcal{V}_{div}(R)
\label{48}
\end{equation}%
where the divergent term is given by%
\begin{equation}
\mathcal{V}_{div}(R)=2\pi \lbrack \frac{R^{2}}{z_{0}^{2}}a+b\log \Big(\frac{%
z_{0}L_{AdS}^{2}}{2R}\Big)],  \label{31}
\end{equation}%
with $a=\frac{3-2L_{AdS}^{2}}{L_{AdS}},\ b=6L_{AdS}^{5}-5L_{AdS}^{3}$.

\begin{figure}[tbp]
\centering
\includegraphics[width=0.24\textwidth]{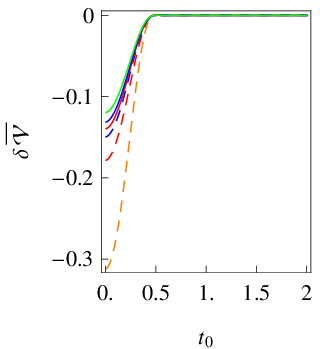} %
\includegraphics[width=0.24\textwidth]{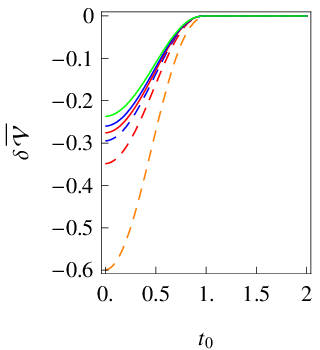} \includegraphics[width=0.24%
\textwidth]{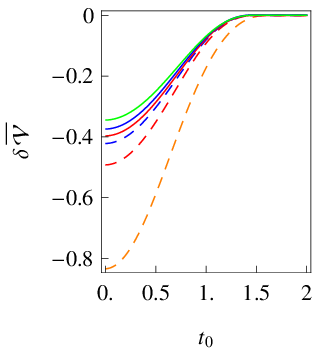} \includegraphics[width=0.24\textwidth]{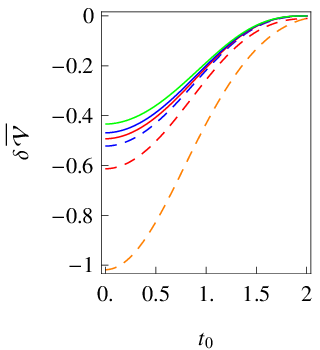}
\caption{$\protect\delta \bar{\mathcal{V}}$ as a function of boundary time $%
t_{0}$ for boundary radii $R$=0.5,~1,~1.5,~2 (from left to right) with
different $\protect\alpha $.}
\label{dyhee}
\end{figure}

\begin{figure}[tbp]
\centering
\includegraphics[width=0.43\textwidth]{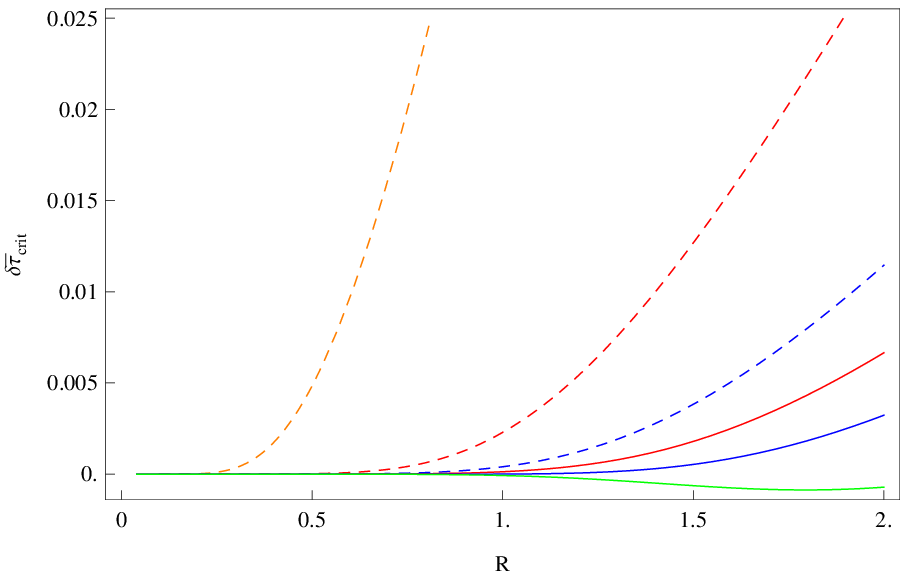} %
\includegraphics[width=0.42\textwidth]{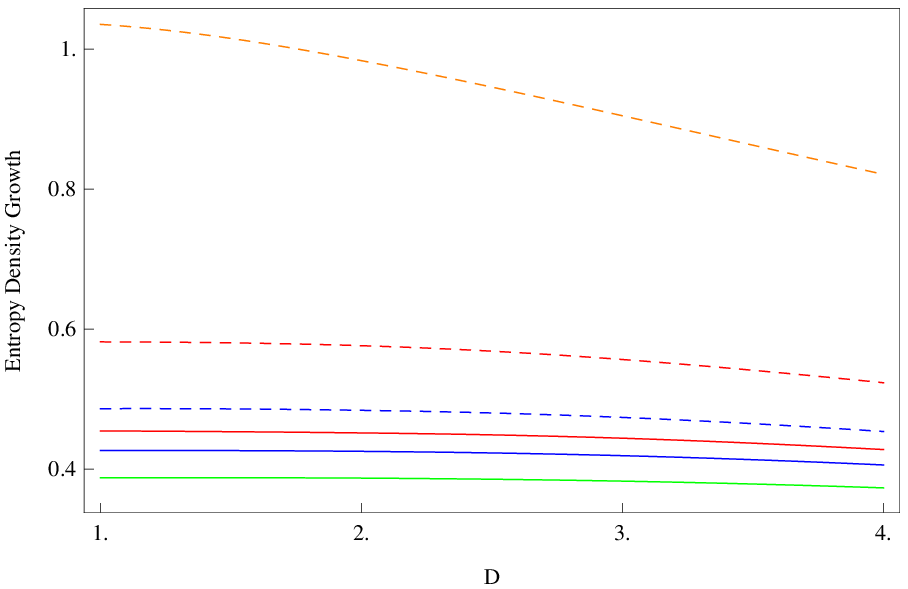}
\caption{(Left) The relative critical thermalization time $\protect\delta
\bar{\protect\tau}_{crit}$ as a function of spatial scale with different $%
\protect\alpha $. (Right) Maximal growth rate of entanglement entropy
density ($\frac{d\protect\delta \bar{\mathcal{V}}}{dt_{0}}$) as a function
of diameter of entangled region. }
\label{timeandrate}
\end{figure}

\begin{figure}[tbp]
\centering
\includegraphics[width=0.43\textwidth]{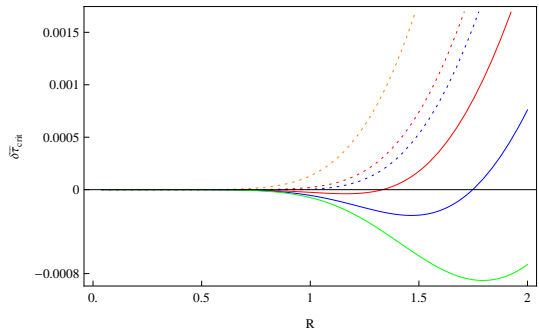}
\caption{The relative critical thermalization time $\protect\delta \bar{%
\protect\tau}_{crit}$ of EE shows certain phase transition with $\protect%
\alpha $=-0.05 (orange-dashed), -0.09 (red-dashed), -0.1 (blue-dashed),
-0.12 (red), -0.15 (blue), -0.19 (green).}
\label{phase}
\end{figure}

In Fig. \ref{dyhee} we plot $\delta \bar{\mathcal{V}}=\frac{\delta \mathcal{V%
}(R,t_{0})-\delta \mathcal{V}_{thermal}(R)}{4\pi R^{3}/3}$ as a function of
boundary time $t_{0}$ for various $R$ and $\alpha $. The relative critical
thermalization time $\delta \bar{\tau}_{crit}$ is also plotted in the left
panel of Fig. \ref{timeandrate}. For small $R$ the thermalization time $\tau
_{crit}$ is linear with $R$. This can be understood since Eqs. (\ref{eq1})
and (\ref{eq2}) also hold for EE with $R_{\ast }=R$. Meanwhile, one may
notice that the relationship between $\delta \bar{\tau}_{crit}$ and $R$
encounters certain \textquotedblleft phase transition\textquotedblright\ at
certain large negative $\alpha $, provided that we take $R$ as an order
parameter{\footnote{%
It is not a new idea to take the boundary separation as an order parameter,
see \cite{order} for instance.}} and $\delta \bar{\tau}_{crit}$ as certain
\textquotedblleft free energy\textquotedblright . To be more clear, see Fig. %
\ref{phase} and one can read the critical coupling constant as $\alpha
\lesssim -0.1$. In particular, note that the $\tau _{crit}$ of EE violates
the causal bound (i.e. $\tau _{crit}\geqslant R$) for the large negative $%
\alpha $ when $R$ is in a certain interval. Also note that the causal bound
is violated for other two probes in Einstein gravity, while not for EE \cite%
{Bala11}.

Now it is time to compare the right panels of Figs. \ref{timeandratetp}, \ref%
{timeandratewl}, and \ref{timeandrate}. One can find that for small boundary
region the growth rate of EE density is nearly volume-independent, but that
is not the case for other two probes.

\section{Conclusion and discussion}

In this note, using the AdS/CFT correspondence and the GB-Vaidya model we
study the thermalization process of the dual boundary field theory. The
two-point functions, the expectation values of circular Wilson loops and the
EE with circular entangling surfaces are studied as the probes. When
boundary separation $R${\ (or }$\ell /2$ for two-point functions) is small,
we observed that the real thermalization times of these observables have the
weak dependence on the GB coupling constant $\alpha $. We also found that
for two-point functions and Wilson loops, the critical thermalization time
has the non-monotonic dependence on $\alpha $. For EE, the critical
thermalization time always increases as $\alpha $ increases. We showed that
the critical thermalization time of all three probes behave like $\tau
_{crit}\sim R$ for small boundary separation $R$ but for large $R$, $\tau
_{crit}$ is not linear with $R$. We also showed that the relationship
between the critical thermalization time of EE and the boundary separation
encounters certain \textquotedblleft phase transition\textquotedblright\ at
large negative $\alpha $ ($\alpha \lesssim -0.1$). Moreover, as in
Einstein's gravity, it's found that maximal growth rate of EE density is
nearly volume-independent for small volumes.

\begin{figure}[tbp]
\centering
\includegraphics[width=0.43\textwidth]{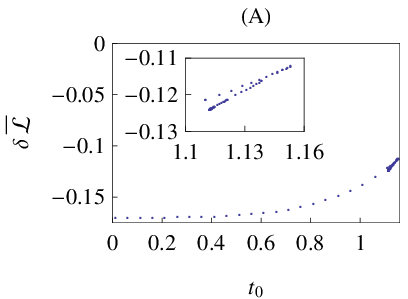} %
\includegraphics[width=0.43\textwidth]{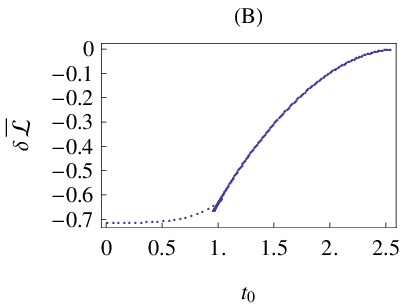} %
\includegraphics[width=0.43\textwidth]{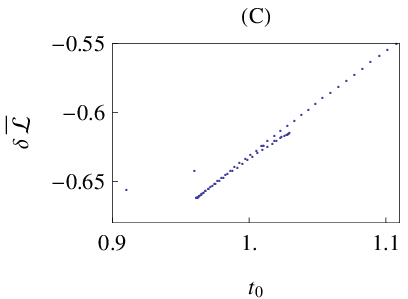} %
\includegraphics[width=0.40\textwidth]{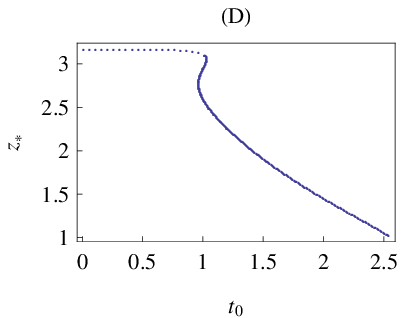}
\caption{(A) The two-point function with $R\simeq 2.157$ and $\protect\alpha %
=-0.19$. Note that here our numerical method is not precise enough to plot
the complete thermalization process. (B) The two-point function with $R=3.2$
and $\protect\alpha =-0.01$. One can see a swallow tail. (C) A zoomed-in
version of the swallow tail. (D) $z_{\ast }$ against $t_{0}$ for $R=3.2$ and
$\protect\alpha =-0.01$.}
\label{st}
\end{figure}

One may wonder if the larger boundary separation could lead to the new
thermalization behavior. Interestingly, we find that there exist different
geodesics with extremal lengths for a range of $t_{0}$. Consequently, the
two-point function behaviors as a multiple-value function, see (A) in Fig. %
\ref{st}. We have found such kind of multiple-value functions when $\alpha $
is negative and $R$ is large enough (For instance, the similar
multiple-value functions appear when $\alpha =-0.1$, $R\simeq 2.403$ and $%
\alpha =-0.05$, $R=2.4$). However, we have not found the similar
multiple-value functions for other two probes with the parameter spaces in
which our numerical method is reliable. Moreover, we note that our numerical
method is not precise enough to plot the complete thermalization process in
(A) of Fig. \ref{st}. Fortunately, when $\alpha =-0.01$, $R=3.2$, we can
plot the complete thermalization process, see (B) of Fig. \ref{st}. It is
very interesting to observe a swallow tail, see (B) and (C) of Fig. \ref{st}
(We have also obtained the complete thermalization process with the similar
swallow tail for several other parameters, for instance, the case with $%
\alpha =-0.03$, $R=2.7$). Comparing (A) and (C) of Fig. \ref{st}, one can
suspect that the multiple-value function in (A) of Fig. \ref{st} may be an
incomplete part of a similar swallow tail. It should be pointed out that the
swallow tail in the two-point function with GB correction is not same as the
swallow tail observed previously in strip EE, where the swallow tail is on
the top of the curve, see for instance Fig. 18 in the second reference of
\cite{Bala11}. However, the swallow tail in Fig. \ref{st} is on the bottom
of the curve. In addition, the previous swallow tail appears near the
thermalization time, but it is not the case for the swallow tail in Fig. \ref%
{st}. The difference between two kinds of swallow tails can also been seen
by comparing the $z_{\ast }$ function of $t_{0}$, see (D) of Fig. \ref{st}
for the two-point function and Fig. 27 in the second reference of \cite%
{Galante1205} for strip EE. The physical meaning of this new kind of swallow
tail deserves to be further studied. Here we only note that the
multi-valuedness of strip EE is avoided by selecting the extremal surface
with the minimal area \cite{Hubeny0705}. However, if we adopt the similar
selection, the two-point function will be not continuous.

\begin{figure}[tbp]
\centering
\includegraphics[width=0.24\textwidth]{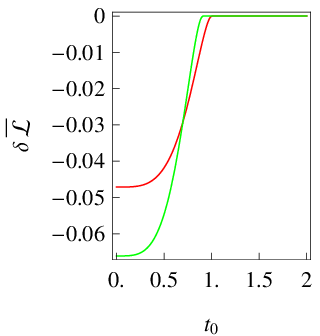} %
\includegraphics[width=0.24\textwidth]{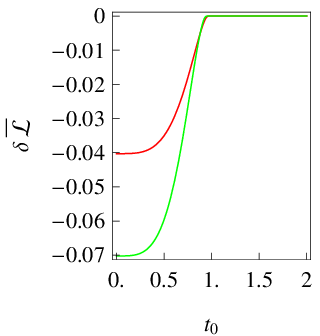}
\caption{Thermalization of the renormalized geodesic lengths using the
metric (2.14) of Ref. {\protect\cite{GBHT}. The green and red lines
correspond to }$\protect\alpha =0.08$ and $-0.1$, respectively. (Left) The
boundary separation $\ell =2$ (see Fig.~2 in \protect\cite{GBHT}). (Right) $%
\ell =2/L_{AdS}$. One can see that the dependence of the thermalization time
on $\protect\alpha $ is suppressed by the scale transformation.}
\label{zxx}
\end{figure}

At last, it should be noted that in Ref. \cite{GBHT}, the authors calculated
the two-point function and rectangular Wilson loop in the field theory dual
to GB gravity. So their work on the two-point function is overlapped with
ours. However, they did not notice the swallow tail which is present outside
the parameters space they scanned. Moreover, they did not show that the
thermalization times only have weak dependence on the GB coupling constant $%
\alpha $. Their results can be changed to ours if one makes a scale
transformation on the boundary separation according to the difference
between our metric and theirs, see Fig. \ref{zxx}.

\begin{acknowledgments}
We thank Y. P. Hu and X. H. Ge for valuable discussions. This work was
supported by National Natural Science Foundation of China (No. 11275120).
\end{acknowledgments}

\end{document}